# Tensile strain induced changes in the optical spectra of SrTiO$_3$ epitaxial thin films

**A. Dejneka[1]\*, M. Tyunina[2], J. Narkilahti[2], J. Levoska[2], D. Chvostova[1], L. Jastrabik[1] and V. A. Trepakov[1,3]**

[1] Institute of Physics AS CR, Na Slovance 2, 182 21 Praha 8, Czech Republic.
[2] Microelectronics and Materials Physics Laboratories, University of Oulu, PL 4500, FI-90014 Oulun Yliopisto, Finland
[3] A. F. Ioffe Physicsol-Technical Institute RAS, 194 021 St-Petersburg, Russia

\*Email: dejneka@fzu.cz

**Abstract:** Effect of biaxial tensile strains on optical functions and band edge transitions of ultra thin epitaxial films was studied using as an example a 13 nm thick SrTiO$_3$ films deposited on KTaO$_3$(100) single-crystal substrates. Optical functions in the 1200 – 200 nm spectral range were determined by spectroscopic ellipsometry technique. It was found that tensile strains result in a shift of the low energy band gap optical transitions to higher energies and decrease the refractive index in the visible region. Comparison of the optical spectra for strained SrTiO$_3$ films and for homoepitaxial strain-free SrTiO$_3$:Cr (0.01 %at.) films deposited on SrTiO$_3$(100) single crystalline substrates showed that this "blue" shift of the band gap could not be related to technological imperfections or to reduced thickness. The observed effect is connected with changes in the lowest conduction and in the top valence bands that are due to increase of the in-plane lattice constant and/or onset of polar phase in the tensile strain-induced ultra-thin epitaxial SrTiO$_3$ films.
PACS numbers: 73.22.-f, 77.84.-s, 78.20.-e

**1. Introduction**
Properties and functional ability of epitaxial films are widely studied nowadays. In this regard thin films of strontium titanate SrTiO$_3$ (STO), a model of the ABO$_3$ and related perovskites, are attracting a special interest. Unique properties provide for various applications of STO in modern nano-electronics and optics: integrated devices, random access memory, capacitors, ultra-thin gate dielectric layers, new generation of the microwave tuneable and electroluminescence elements, nanoscale bistable resistivity switchers, etc (see, e.g. [1-7]). So far, studies of STO thin films have been focused mainly on the structure and dielectric properties. Only few studies were devoted to basic optical characteristics of STO thin films, such as refractive index and band edge optical transitions (e.g. [8-17]) controlling principal photo-electronic properties. Even so, these studied have been performed with thin films of different quality, polycrystalline (or different degree of crystallinity), and/or amorphous films, deposited by various techniques on various substrates with pronounced films-substrate interfacial layers, imperfections etc. The obtained contradictory results have been considered disputable and explicit intrinsic properties inherent to high-quality epitaxial STO thin films have not been presented to date. To summarize, the structural imperfections, presence of amorphous interfaces and oxygen vacancies in most cases call for a distinct reduction of the refractive index in the visible region.
In most cases, the band-gap energies of poor-crystallized films with pronounced interlayer effects appear to be larger than those of single crystals (e.g. [8, 12, 15]), although it is not always the case [9], The band-gap of highly crystallized films appears to be at least comparable to that of single crystals [10,14,15]. To our knowledge, optics of sufficiently high nanometric quality of STO films was studied in [13] (polycrystalline films on silica targets) and [14] (homoepitaxial, metallic, oxygen-deficient STO films on STO substrates) in which no conspicuous band edge shift has been found.

Recently, the fabrication and studies of 2D clamped heteroepitaxial strained STO films have attracted a special interest. Through the series of theoretical and experimental works (e.g. [3,19-21]) it is believed today that polarization response and tuneability of such films are influenced crucially by misfit strains. In STO films with biaxial in-plane tensile strains (films grown on tensile straining substrates), the onset of ferroelectricity with the transition temperature $T_C$ up to room temperatures (RT) has been demonstrated [19, 21]. This phenomenon is one of the leading research topics today and stimulates active technological and physical studies of the strained epitaxial films. However, reports on optical properties and band gap energies of strained thin STO films are virtually absent.

The present work deals with visible and near-UV optical spectral ellipsometric studies of the complex optical dielectric function of STO films epitaxially grown on KTaO$_3$ (KTO) substrate (heteroepitaxial STO/KTO films). Due to large mismatch between lattice parameters of cubic perovskite STO ($a_{STO}$ = 3.905 Å) and KTO ($a_{KTO}$ = 3.989 Å), biaxial in-plane tensile strain as large as 2.2 % is expected in STO/KTO. To sustain such strain, films with thickness of only several nanometres should be grown [22]. For comparison, homoepitaxial strain-free films of SrTiO:Cr(0.01%at.) on STO(001) single crystals (STO/STO films) were sintered and studied too. Such a small quantity of Cr admixture does not practically influence band edge and interband transitions, but even very small change of the refractive index made by Cr impurities makes it possible to apply the spectral ellipsometry technique more successfully. Studies of homoepitaxial strain free STO/STO films allowed us to take into account possible interface and thickness effects. Additionally, as a reference and for reliability of the received data, optical properties of STO single-crystals were measured too.

## 2. Experimental
*2.1. Epitaxial SrTiO$_3$ films*

Top seed solution-grown KTO single crystals from University of Osnabrueck were used as substrates for heteroepitaxial STO films. Substrates have been prepared as thin <100> oriented plates with optically polished surfaces.

STO and STO:Cr films with thickness 13 nm were deposited by *in situ* by pulsed laser deposition on KTO(001) and STO(001) substrates, as described in [23]. For fabrication of STO:Cr films, the Verneuil grown STO:Cr (0.01% at.) single crystals were used as a target. Room-temperature x-ray diffraction (XRD) analysis of STO/KTO revealed that STO films were perovskite, epitaxial, with (001) planes parallel to the substrate surface, and with the epitaxial relationship STO[100](001)||KTO[100](001). The out-of-plane and in-plane lattice parameters were found to be $c \approx$ 3.864 Å and $a \approx$ 3.987 Å, respectively. The biaxial in-plane tensile strain was about 2.1 %, being consistent with the expected misfit strain of 2.2 %. STO/STO film exhibited XRD patterns with the Bragg peaks completely coinciding with those of the substrate suggesting good homoepitaxial growth. More details on films deposition and microstructure have been published in [24].

*2.2. Spectroscopic ellipsometry*

To study in-plane optical properties of STO films, ellipsometric spectra were collected using a variable-angle spectroscopic ellipsometer J. A. Woollam operating in a range of photon energies $E$ = 1 – 6.2 eV (wavelength $\lambda \sim$ 1200- 200 nm). The main ellipsometric angles $\psi$ and $\Delta$ were measured in the reflection mode at angle of incidence 65-75°. To eliminate depolarizing effects of back-surface reflection from the substrate, the back surface of the substrates was mechanically roughened. The ellipsometric spectra were analyzed with the WVASE32 software package considering a stack of semi-infinite substrate, STO film, surface roughness, and ambient air. To improve the fitting procedure and reduce the number of fitting parameters, all substrates were measured independently and their calculated optical constants were used in further STO films analysis as fixed parameters. Resulting surface roughness of KTO and STO substrates was ~ 2 nm or less. The perturbation of the ellipsometric data was mathematically removed by assuming that the surface roughness could be approximated as a Bruggeman effective medium [19]. For spectral dependences of optical constants the generalized multi-oscillator model was used. The thickness of STO films calculated from ellipsometric data was in a good agreement with XRD measurements. Resulting roughness of KTO surface roughness was less then 2 nm.

## 3. Results
*3.1 SrTiO$_3$ Crystals.*

*Tensile strained SrTiO$_3$ films*

Figure 1 shows the resulting complex index of refraction ($n^*=n+ik$ ) as determined from the spectroscopic ellipsometry measurements for the bulk STO single-crystal specimen. The magnitude of the real part of the index of refraction at 2.06 eV (600 nm) is 2.38, which agrees well with the reference [25] and ellipsometric data of *Zollner et al* [13]. It is seen that refraction in visible region is controlled by two refractive index maxima located at 3.88eV (320 nm), 4.10 eV (302 nm) and 4.66 eV (266 nm).

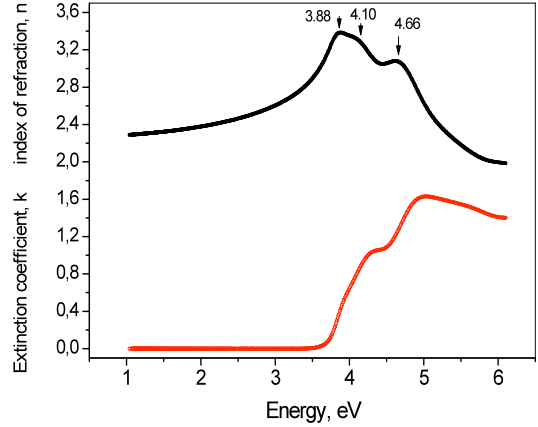

Figure 1. Complex index of refraction for bulk STO extracted from spectroscopic ellipsometry.

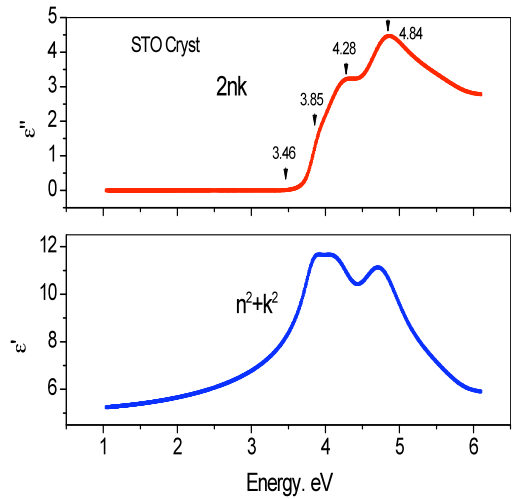

Figure 2. Dielectric functions of SrTiO$_3$ crystal.

Figure 2 presents real and imaginary part of the dielectric constant for STO bulk crystals determined from spectroscopic ellipsometry. Obtained $\varepsilon'$ and $\varepsilon''$ spectra appear to be very close to those reported by *Zollner et al* [13] and are even better resolved than those presented by *Cardona* [26] for STO bulk crystals. Only the pronounced $\varepsilon''(h\nu)$ maximum at 3.85 eV observed by *Zollner* corresponds to a shoulder in our experiment; however, we resolved two maxima at 4.28 and 4.84 eV. Conventional formula $\alpha = (4\pi k)/\lambda$ allows us to obtain spectrum of the absorption coefficient, which is very close to those reported by *Cardona* [26].

Although spectral ellipsometry does not allow sufficiently accurate determination of the absorption coefficient magnitudes below 3.2 eV, where absorption is weak, our measurements detected the obvious presence of a long exponential Urbach-type absorption tail $\alpha=\alpha_0 \exp [-\gamma(E_0-E)]$ near the absorption edge of STO crystals. In agreement with [13], this exponential tail spreads up to the point where absorption coefficient rises to ~ 100 cm$^{-1}$ and at higher energies is goes over to spectral region of indirect gap transitions with $E_i$ = 3.29 eV, as determined from the intercept of a linear fit to a plot of $\alpha^{1/2}$ versus energy. The obtained indirect character of band gap in STO and the magnitude of indirect band gap agree well with those reported for STO in [10,13,27-29]. In agreement with experiments and calculations [13,28-32] it corresponds to $R_{15'} \rightarrow \Gamma_{25'}$ indirect transitions between the lowest CB and upper VB. Besides, our analysis shows a possibility of another indirect transition with energy 3.64 eV, which agrees well with the results [33] and can be attributed to $M_{51} \rightarrow \Gamma_{52'}$ gap. At higher energies, up to about 5 eV, the absorption coefficient obeys the relationship $\alpha \cdot h\nu^2 \sim h\nu - E_d$, evidencing direct optical transitions with energies 3.81 eV, 3.9 eV and 4.34 eV. They can be tentatively attributed to direct gaps between Γ (as in [13, 28, 33]), X (as in [26, 33]) and X or Γ (as in [13]) points of the Brillouin zone respectively. The broad maximum at ~ 5 eV seems to be similar to the A$_3$ one reported by *Cardona* [26]. It can be caused by direct interband optical transitions in X or M points of the Brillouin zone. Hence, we see that our spectral ellipsometry is correct and can be used for further studies of STO films.

### 3.2. SrTiO$_3$/KTaO$_3$ Films

Figure 3 and 4 present complex index of refraction and absorption coefficient spectra calculated as $\alpha = (4\pi k)/\lambda$ obtained for STO/KTO film in comparison with those for STO bulk crystals.

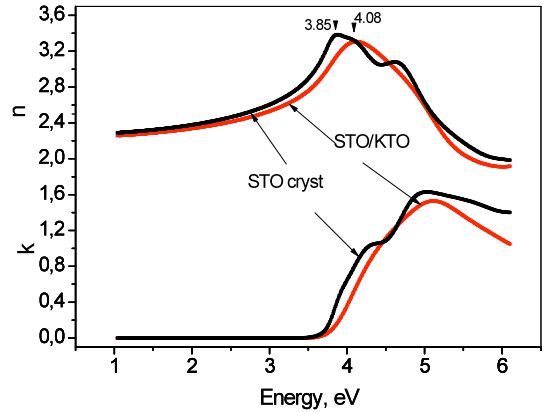

Figure 3. Complex index of refraction for SrTiO$_3$ crystals and STO/KTO film.

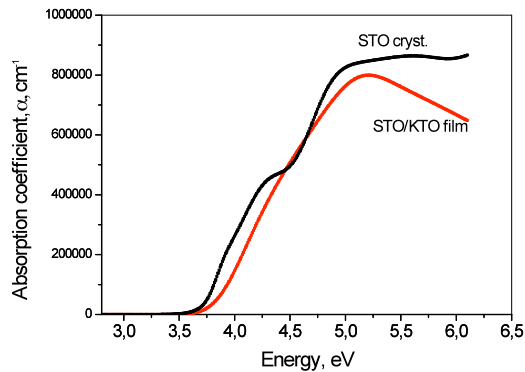

Figure 4. Spectral dependences of the absorption coefficient for STO bulk and STO/KTO film calculated as $\alpha = (4\pi k)/\lambda$

Three obvious effects can be distinguished right away for STO/KTO film in comparison with STO bulk crystals: *i*) the magnitude of the refractive index at 600 nm (2.06 eV) is 2.36, which in contrast to *epitaxial films* reported in [14] is only slightly smaller than that in bulk crystals, evidencing perfect quality of our films; *ii*) the band edge is shifted to higher energies, *iii*) the spectrum of inter-band transitions looks smoothed. It is clearly seen that the reduction of refractive index in visible region is mainly caused by a shift of the $n(h\nu)$ spectral maximum in the ~ 4 eV region to higher energies. At the same time, in STO/KTO films this maximum is located at 4.08 eV, close to the A$_1$ maximum reported by *Cardona* [26] and the shoulder in our experiment for STO bulk crystals. Taking into account the spectral dependence of the extinction coefficient k, such transformation can be connected to absence or strong suppression in STO/KTO films of the $n(h\nu)$ maximum at 3.85 eV present in STO bulk crystals.

As in STO crystals, the pronounced long exponential Urbach-type absorption tail (one or two) presents in tensile strained STO/KTO films substituting the onset of indirect optical transitions at high energies. The corresponding $\alpha^{1/2}(h\nu)$ plot results in indirect band gap magnitude of 3.82 eV. At higher energies the intense absorption obeys the relation $(\alpha h\nu)^2 \sim h\nu - E_d$ evidencing two direct optical transitions with energy 4.16 and 4.32 eV. At last, the strong maximum of absorption coefficient at ~ 5.1 eV can be attributed to direct optical transition, too.

*3.3. SrTiO₃/SrTiO₃ Films*
As it was pointed above, the properties of epitaxial ferroelectric thin films can be generally affected by strain, presence of surface and interface layers, internal and random electrical fields, stresses, technological imperfections, oxygen vacancies, etc.. To evaluate qualitatively the possible effect of strain, optical properties of homoepitaxial STO:Cr /STO film were analyzed.

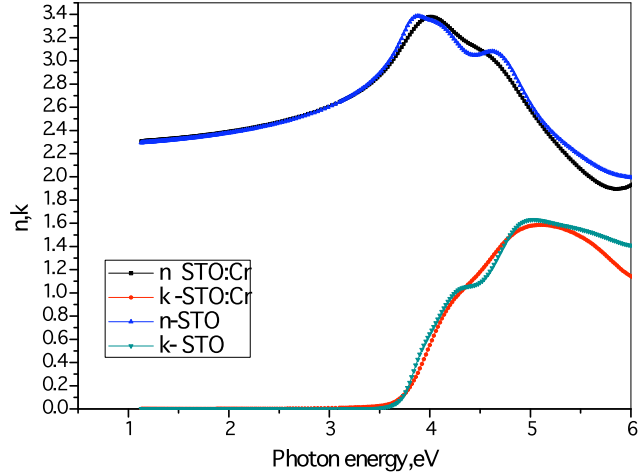

Figure 5. Refractive index and extinction coefficient spectra of STO crystal and STO:Cr /STO thin film.

As seen from figure 5, both STO:Cr /STO film and STO bulk crystal have the same magnitude of refractive index in the visible region, virtually identical magnitudes of optical band edge and very close character of near band-edge interband optical transitions. However, some spectra smoothing and transformations resemble those for STO/KTO films. The presence of interface and small thickness itself might be probably responsible for the smoothing, leading, e. g., to a broadening of critical points of Brillouin zone. The actual mechanism of the smoothing is unclear, but it is not subject of this study. Our attention was paid to the comparison of optical spectra for heteroepitaxial tensile strained STO/KTO films with those for unstrained homoepitaxial STO:Cr/STO films, which allowed us to ascribe the dramatic changes (blue shift of absorption edge) in the spectra of STO/KTO films mainly to the presence of biaxial tensile strain.

**4. Discussion.**
At first, let us to consider the possible nature of the observed tensile strain effect using crystal-optics approach. Homogeneous biaxial tensile strain of STO film reduces its symmetry from cubic m3m to tetragonal 4/mm. In such case the in-plane optics non-diagonal contribution of $T_{zz}$ stresses and related deformations can be neglected. The changes of in-plane refractive index of such film can be written as:
$$b = b_p + b_{pt} \qquad (1)$$
where $b$ is the change of refractive index, $b_p$ – describes photoelastic contribution controlled by changes in interband transitions (in the simplest case - band-edge shifts) due to in-plane lattice constant expansion, $b_{pt}$ – describes possible contribution of the phase transition. In most cases, the photoelastic contribution is negative, $b_p = dn/d\alpha < 0$ ($\alpha$ is a lattice constant), because actual tensile strain decreases the in-plane density and so tends to diminish the refractive index with rising strain [34]. Following [34] the photoelastic contribution in tensile strained STO films can be written as:
$$\Delta n = -\frac{n^2}{2}\left[(p_{11} + p_{12})\Delta\alpha_{1,2} - p_{13}\Delta\alpha_3\right]$$
(2)
where $p_{ij}$ is the photoelastic tensor, $p_{22}$ and $p_{12}$ are in-plane, and $p_{13}$ out-of-plane components. Using magnitudes of the photoelastic tensors from [35]: $p_{11} = 0.19$ and $p_{12} = 0.044$ gives $p_{11} + 2p_{12} = -0.10$. Because of this and positive $\Delta a > 0$, the photoelastic contribution appears to be positive, which means a shift of the band edge to lower energies. So, the predictions of crystal-optic photoelastic analysis contradict our observation.

Now, let us try to connect the observed blue shift of the STO/KTO films band gap with the peculiar STO electronic band structure and its changes due to the increase of the lattice constant. According to the theoretical calculations (e.g. [30-32,13,28,36]), hybridization manifest itself very slightly in the structure of the upper valence band (VB) and the lowest conduction band (CB). The upper VB consists of 2p atomic orbitals with a small admixture of Ti and Sr atoms (figure 6). The band is quite flat at M or R points of the Brillouin zone (BZ) and is almost perfectly flat between these points. The bottom of the lowest CB is constructed essentially of Ti 3d $t_{2g}$ and $e_g$ states. Orbital contribution from Sr atoms is negligible in this energy range and Sr 4d $t_{2g}$ and $e_g$ states make a significant contribution to the DOS only at much higher energies. In such situation, the band scheme transformation and $E_g$ dependence versus lattice constant, presented in figure 7, describes the band gap evolution in the 2D tensile strained epitaxial STO/KTO films. This picture is also consistent with the following quantum chemical bonding analysis. According to Harrison [37], in perovskite-like materials such as STO, the dependence of the band gap versus interatomic distance can be found in the following way. The energy of Ti 3d states of the lowest CB can be expressed as $\Gamma_1 = \varepsilon_d$ whereas the energy of the O 2p states of the upper VB can be written as:

$$\Gamma_2 = \varepsilon_p + 2\sqrt{2}(V_{pp\sigma} + V_{pp\pi}), \qquad (3)$$

where $V_{pp\sigma}$ and $V_{pp\pi}$ - are the $\sigma$ and $\pi$ bonding energies of neighbouring O *2p* states.

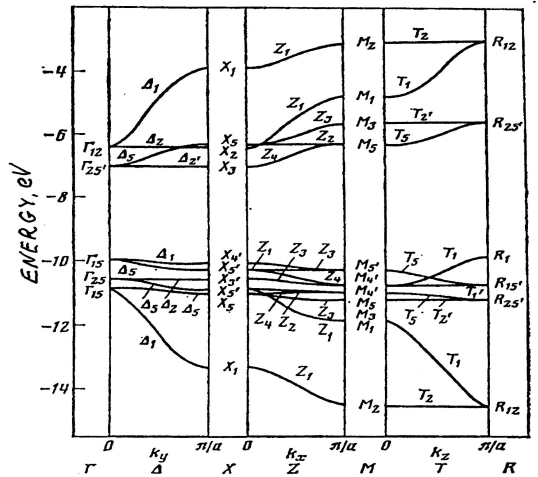

Figure 6. Band structure of SrTiO3 in cubic phase [30]

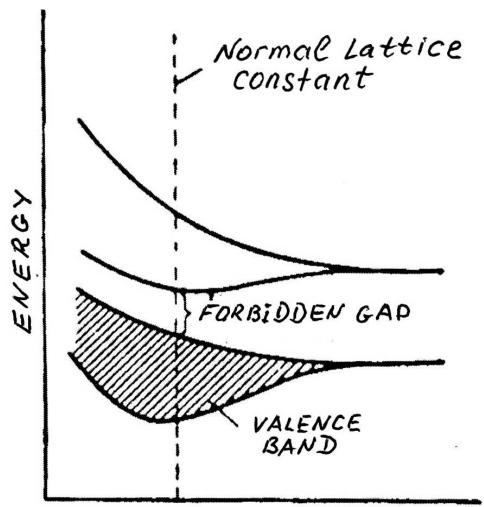

Figure 7. Dependence of the electronic energies (allowed bands) versus interatomic distances.

The sum of $V_{pp\sigma}$ and $V_{pp\pi}$ is:
$$V_{pp\sigma} + V_{pp\pi} = (\eta_{pp\sigma} + \eta_{pp\pi})(\hbar^2/m)d^{-n}, \qquad (4)$$
where $\eta_{pp\sigma}$, $\eta_{pp\pi}$ and $n$ $(n > 0)$ are dimensionless coefficients, $\hbar$ is Planck's constant, $m$ is electron mass, and $d$ is the distance between the neighbouring oxygen atoms. Therefore, for $E_g$ we come to formula:
$$\Gamma_1 - \Gamma_2 = \varepsilon_d - \varepsilon_p + 2\sqrt{2}(\eta_{pp\sigma} + \eta_{pp\pi})(\hbar^2/m)d^{-n}, \qquad (5)$$

from which it is seen that the increase of the lattice constant *d* enhances the $E_g$ magnitude. To put it simply, this effect is caused by a decrease of the orbital overlap integrals and, consequently, a decrease of the lowest CB and upper VB widths and $E_g$ enhancement, as shown in figure 7. However, the obtained result should to be considered cautiously because the molecular approach does not take into account crystal field effects.

It is remarkable that very similar effect had been observed recently for STO nanopowders [38, 39]. It was found that decreasing particle size in STO nanopowders is accompanied by pronounced increase of the lattice constant. Expressing the dependence of the band gap versus the change of the lattice constant $\Delta\alpha$ as $E_g = E_{g0} + k\Delta\alpha$, where $E_{g0}$ is the band gap energy for unperturbed STO, and taking more reliable data from [39] $\alpha$ = 3.910 Å and $E_g$ = 3.38 eV for 20 nm nanopowder, we obtain $k \sim 6.5$ eV/Å, whereas similar estimate for strained STO/KTO films gives $k \sim 18$ eV/Å. It seems that the similarity is not random and it follows from the considered mechanism that the pronounced blue shift of interband optical transitions in STO/KTO tensile strained films with respect to STO crystals (and/or homoepitaxial STO/STO films) mainly concerns band edge transitions, whereas the energy of absorption maximum at ~ 5 eV changes very little (figure 4). Hence, the explanation seems to be found. However, *ab initio* theoretical estimates [40] made on our request using the B3LYP hybrid exchange technique as it had been done in [36], have shown that the energies of all Γ-Γ, X-X, M-M, R-R, X-Γ, M-Γ and R-Γ optical gaps decrease with increasing lattice constant.

Thus, we believe that the mechanism directly connected to blue shift of $E_g$ in the tensile strained STO/KTO films describes at least a part of contributions to the discussed phenomenon, or that the description of the band structure for ultra-thin tensile strained epitaxial STO films calls for special theoretical consideration distinct from that available for crystals.

At the same time, decreasing of the refractive index in the visible region and the "blue shift" of the optical band gap in tensile strained STO/KTO film can be understood as indirect manifestation of the polar phase transition from the tetragonal phase to ferroelectric polar phase induced by non relaxed tensile strains in ultra-thin STO/KTO films as it has been predicted in [19]. Our experimental results on STO/KTO films remarkably have something in common with the results of theoretical [41, 42] and experimental studies of the tetragonal phase transition effect in the band edge optics performed for related BaTiO₃ (BTO) crystals [33]. In these studies it was clearly demonstrated that the cubic-tetragonal ferroelectric phase transition in BTO is accompanied by splitting and distinct shifts of the direct Γ→Γ band edge transitions to higher energies for the light with electrical vector polarized along the spontaneous polarization $P_S$ (i.e. along the direction of lattice constant increase), whereas considerably smaller changes were found for the light polarized normal to this direction.

## 5. Conclusions

In this work we reported the first studies of the tensile strain effect in the optical constant and optical band gap transitions of ultra-thin strain-free 13 nm thick homoepitaxial STO films grown on STO substrates and 2D-clamped in plane tensile strained heteroepitaxial STO films of the same thickness deposited on KTO substrates. The characteristic refractive index magnitudes, band gap energies and the energies of main near band-gap optical transitions for in plane polarized light have been determined. The magnitude of the refractive index and optical spectra of homoepitaxial STO/STO films appeared to be nearly the same as those for STO bulk crystals evidencing perfect films crystallinity. Compared to STO crystal and homoepitaxial STO films, heteroepitaxial tensile strained STO/KTO films exhibited a pronounced shift of the band edge and near-edge interband transitions to higher energies, and a reduction of refractive index in visible spectral region. It was clearly shown that these effects could not be related to technological imperfections or to reduced thickness. Detailed treatment of the experimental results allowed us to suggest that the reduction of the refractive index and the shift of the band edge optical transitions to higher energies observed in STO/KTO films are caused by increasing of the in-plane lattice constant immediately and onset of polar phase inherent in tensile strain-induced ultra-thin epitaxial SrTiO₃ films.

## 6. Acknowledgements
Authors thank to E. Kotomin and S. Piskunov for fruitful discussion and theoretical computing. This work was supported by Grants 1M06002 of MSMT ČR, KAN 301370701, and AV0Z10100522 of AV ČR, GACR 202/08/1009, and Academy of Finland (No 118250), Sc. Sch.-2628.2008.2, and Program of Presidium RAN "Quantum physics of condensed matter".